\documentclass[12pt,letterpaper]{article}
\usepackage[a4paper, total={7in, 10in}]{geometry}

\usepackage{graphicx}
\usepackage{helvet}
\usepackage{authblk}
\usepackage{hyperref}
\usepackage{amsmath} 
\usepackage{amssymb} 
\usepackage{orcidlink} 
\usepackage{siunitx}

\usepackage[super,comma,sort&compress]  
{natbib}
\usepackage[toc,acronym]{glossaries}

\makeglossaries
\newacronym{AP}{AP}{Action Potential}
\newacronym{AFM}{AFM}{Atomic Force Microscope}
\newacronym{APTD}{APTD}{Action Potential's Time Derivative}
\newacronym{ATP}{ATP}{Adenozine Triphosphate}
\newacronym{CNS}{CNS}{Central Nervous System}
\newacronym{EM}{EM}{electromagnetic}
\newacronym{GHK}{GHK}{Goldman-Hodgkin-Katz}
\newacronym{HBP}{HBP}{Human Brain Project}
\newacronym{HH}{HH}{Hodgkin and Huxley}
\newacronym{HW}{HW}{hardware}
\newacronym{PID}{PID}{Proportional-Integral-Derivative} 
\newacronym{PSP}{PSP}{Post-Synaptic Potential}
\newacronym{AIS}{AIS}{Axon Initial Segment}
\newacronym{AI}{AI}{Artificial Intelligence}
\newacronym{AIMC}{AIMC}{Analog In-Memory Computing}
\usepackage[
]{changes}
\definechangesauthor[name=V\'egh, color=blue]{VJ}
\definechangesauthor[name=VJA, color=violet]{VJA}

\makeatletter
\renewcommand{\maketitle}{\bgroup\setlength{\parindent}{0pt}
\begin{flushleft}
  \textbf{\@title}
  
  \@author
\end{flushleft}\egroup}
\makeatother

\title{Is $K^+$ current in Action Potential real?}
\date{}

\author[1,\orcidlink{0000-0002-3247-7810}]{János Végh}

\affil{Kalimános BT, Hungary, Debrecen}

\affil[*]{Correspondence: Vegh.Janos@gmail.com}

\begin{document}

\maketitle
	
\section{SUMMARY}

The correct description of ion traffic during an Action Potential in neurons has fundamental importance from 
its theoretical description to practical clinical applications.
The classical \gls{HH} theory hypothesized that a delayed $K^+$ current causes the hyperpolarization of the \gls{AP}.
The paper shows that the membrane's capacitive current was misinterpreted as $K^+$ current, and that the direct evidence for that current is an experimental artifact, both arising from an incomplete understanding of the electrical model. After introducing the \gls{AIS} (discovered decades after constructing the classical model) and using
the correct cross-disciplinary physical laws for living nature (instead of using laws valid for inanimate nature), one can construct the correct
model for neuronal operation, which provides the perfect 
description of neuronal operation, including the Action Potential, in full accordance with the laws of science.

\section{KEYWORDS}

%%%  Include up to 10 keywords, separated by commas. 
%%%  Keywords entered in EM are not carried over; only 
%%%  keywords included in the main text will be used 
%%%  in the final article metadata. Please note that 
%%%  for some journals, keywords are chosen by the editors.

neuron model, net electric neuron model, heat absorption, potassium current

%\pagebreak
%\added{
%	The table of contents is present only for the comfort of reviewers}
%\tableofcontents

\section{Introduction}

Emitting Action Potentials is a fundamental aspect of neuronal operation.
\gls{HH} hypothesized that neurons operate with electrical currents and 
suggested the idea~\cite{HodgkinHuxley:1952} that the neuronal operation
is more or less reminiscent of an electric oscillator. Given that the parallel $RC$ circuit they hypothesized cannot produce an output voltage with opposite sign, they had to introduce
the concept of a delayed current in the opposite direction. Even, they demonstrated that $K^+$ current flows in the second half 
of the \gls{AP}. The net electrical model led immediately to contradiction with the empirical evidence that, 
as Hodgkin wrote~\cite{HodgkinConduction:1964}, "Hill and his colleagues found~\cite{HeatProductionNeuron:1958} that an initial phase [of the action potential] was followed by one of
heat absorption. [...] a net cooling on open-circuit was totally unexpected
and has so far received no satisfactory explanation."
Not to mention the mechanical, optical, etc. effects~\cite{MechanicalWaves:2015}; accompanied by \gls{AP}. 
Neglecting the \gls{AIS} and sticking to \gls{AIS}-less model 
led to clue-less ad-hoc assumptions and desperate claims such as 
"The sodium influx associated with
depolarization is exothermic  [...] the capacitive energy stored in the membrane is
released as heat. However, the efflux of $K^+$  ions is endothermic [...] \textit{converting heat in the environment
into capacitive energy at the membrane}."~\cite{MechanicalWaves:2015}
(That anti-dissipation effect would be something new for physics and could convert the harmful global warming to an
inexhaustible energy source for humankind.)
The correct physical explanation can be derived using the 
unified model of neuronal operation~\cite{Vegh2025ThermodynamicModel:2026,VeghUnifiedCrossdisciplinaryModel:2026}: under the mechanical pressure
caused by the mutual repulsion of the ions, the electrolyte
is compressed and expanded.

Fortunately, the correct neuronal model can provide
reasonable explanation also for that mystic current, explain its origin, and interpret the artifact in the measurement~\cite{HodgkinHuxleyCurrent:1952} that underpinned the existence of the $K^+$ current. In section~\ref{sec:ActionPotential} we introduce the correct 
physics and mathematics of the \gls{AP}. 
We show that the current believed to be a delayed $K^+$ current is actually the capacitive current of the 
condenser in the serial $RC$ circuit, where the $R$ is
the resistance of the \gls{AIS} (instead of the parallel resistance of the ion channels in the membrane's wall), which is consequently omitted from the classical \gls{HH} model.
In section~\ref{sec:OutwardK} we interpret physically the
outward current, which is really an outward  $K^+$ current
through the membrane's wall, but exists only when clamping is used. That is, hypothesizing the non-existent $K^+$ current
was required due to assuming the
wrong oscillator model (omitting the \gls{AIS} that was not yet discovered), and it was mis-identified as a genuine $K^+$ current that does not exist during a native \gls{AP}. 
Our discussion automatically challenges the role of "delayed $K^+$ channels" in producing \gls{AP}, and in general, 
the role of protein-controlled ion channels in neuronal operation.

\section{The Action Potential\label{sec:ActionPotential}}

There is a large number of ion channels between the neuron's internal and external worlds. In addition to the gated $Na^+$ ion channels, a smaller number of non-gated ion channels are distributed over the surface of the membrane, and the overwhelming majority of non-gated ion channels are concentrated in the \gls{AIS}. Given that the conductance of \gls{AIS} is about 
two orders of magnitude higher than that of the distributed ion channels~\cite{ActionPotentialGenerationNatrium:2008,AIS_Updated_Viewpoint:2018}, we approximate that the latter plays a role only in the resting state, while the former plays a role only in the transient state.

Introducing fixed-voltage batteries (in the picture of "equivalent electrical circuit") for illustrative purposes
might be helpful for beginners, but it prevents discussing 
voltage changes in the neuron during producing an \gls{AP}, including the ones illustrated in our figures.

\subsection{The physical process\label{sec:AP-PhysicalProcess}}

In the resting state, we have a condenser with some low-conductance distributed resistors (the resting ion channels) across its plate, and a very low-intensity distributed current (not identical to the famous "leakage current") flows. Due to the slightly fluctuating membrane potential, a current through the always-open channels flows on the dendritic surface
and the resting channels' conductivity represents a drain with 
sufficient transmission. The resting current practically does not 
reach the \gls{AIS}: the ion channels along the current's path
"shunt" the current.
This is a static state that classical neuroscience assumes: some static current in and some static current out; no significant gradient.
\textit{In this state}, an almost correct model is a condenser with a
\textit{parallelly} switched resistor; an integrator-type  $RC$ circuit
(although when approaching the threshold voltage, the \gls{AIS} plays some role). The smaller gradient changes, such as subthreshold excitations, produce "mini-\gls{AP} waves"~\cite{NeuralEnergyConsumption:2017}, providing a direct experimental proof that in this state the parallel $RC$ circuit is not correct.
%The small perturbations are counterbalanced, but their
%amount does not exceed a predefined threshold.

At the beginning of the transient state, the process variable
(the membrane's potential) exceeds the threshold. That event acts
as a fast trigger signal and a new physical process appears. The amount of charged particles (and so:
the membrane's voltage) suddenly increases; the created electrochemical  gradient
drives a current. Notice that in addition to the chemical and electrical gradient, also an elastic gradient due to the 
movement of the elastic membrane affects ions' movement, and even dominates the phenomenon. 
The classical model considers only the electrical gradients,
and even that gradient in a very limited form.
When $Na^+$ ions rush into the intracellular layer, they roughly increase the overall concentration and the potential in that thin layer.
All other ions, including $K^+$, also feel a driving force. The targeted membrane potential and concentration in the thin layer near the membrane are set electrochemically,
and the local driving gradients in the bulk may behave unexpectedly during the transient period.
The actual voltage gradient may temporarily reverse the direction of the chemical gradients.

Our results align with the observation (see caption of Fig.~11.22 in~\cite{MolecularBiology:2002}), that \textit{the significant processes
	occur in a thin layer of the electrolyte proximal to the membrane surface}.
The amount of unbalanced ions is in the range of $10^7$,
and so is the amount of rush-in ions. In addition, those 
ions on the high-concentration side rush into the 
low-concentration side and cause a significant change in the 
membrane potential (and concentration) on both sides of the membrane. Their absolute amount is small compared to the total number of ions in the cell,
but it is significant compared to the number of unbalanced ions in that layer.
However, the layer itself can also be modeled
as having just a few ions under their mutual repulsion on the surface
or a few atomic layers on top of each other, depending on the concentration.

Suppose that at the beginning of an \gls{AP} a large amount of $Na^+$ ions are transferred from the extracellular to the intracellular side. (The concentrations and voltages change with time, so we provide values corresponding to assuming instant interaction, but keeping in mind that the interactions are slow.)
 In that case,  the $K^+/Na^+$ concentrations change from $145/15$ to $45/115$ and the 
corresponding thermodynamic voltage contribution changes from $+61\ [mV]$ to $-25\ [mV]$ (i.e., altogether a $86\ [mV]$ sudden increase in the value of the membrane potential). For the $Na^+$ ions, the resultant potential changes from $+4\ [mV]$ to $-83\ [mV]$. That means when the \gls{AP} begins, the "Na-K pump" stops (if the resulting potential provides the driving force for the exchange pump). The only way for the neuron to remove the excess  $Na^+$ ions is to generate a current toward the \gls{AIS} (where the other end of the ion channels remained at the extracellular potential), until the $Na^+$-specific driving force disappears. The $K^+$-specific driving force, changes from $-90\ [mV]$ to  $-176\ [mV]$, i.e., the $K^+$
outflow gets more intensive, misleading researchers into believing that it causes
the observed hyperpolarization. However, as  Fig.~3 in~\cite{NeuralEnergyConsumption:2017} demonstrates, it occurs instantly, not with a delay; furthermore, the "leakage current" and the stimulus current are negligible. The low extracellular $K^+$ concentration plus the 
low number of channels in the membrane's wall
do not enable a significant increase in the intracellular $K^+$. 
(Given that complex changes occur, including changes in the electrical potential that also change the concentrations, different waves start; the statement is not strictly valid.)

\subsection{The electrical model}
\begin{figure}[]
	\includegraphics[scale=1]{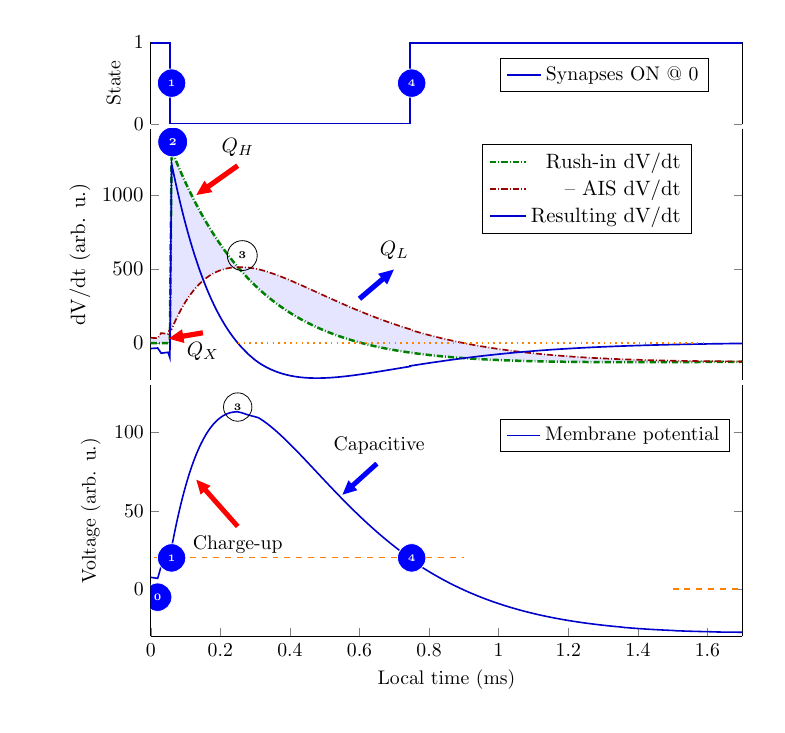}
	
	\caption{The physical processes describing the membrane's operation\label{fig:PhysicalProcessesMembrane}. The rush-in $Na^+$ ions instantly increase the membrane's charge, and the membrane's capacity discharges (producing an exponentially decaying voltage derivative). The ions are created at different positions on the membrane, so they take different times to reach the \gls{AIS}, where the current produces a peak in the voltage derivative. The resulting voltage derivative, the sum of the two derivatives (the \gls{AIS} current is outward), drives the oscillator. Its integration produces the membrane potential. When the membrane potential crosses the threshold, it switches the synaptic currents on or off.}
	
\end{figure}

In the electrical view, the process can be modeled as follows: a sudden voltage gradient  (a sudden jump, followed by a discharge; see Figure~\ref{fig:PhysicalProcessesMembrane} middle inset) appears on the membrane (as a discrete capacitor), and a current flows out (see Figure~\ref{fig:PhysicalProcessesMembrane} bottom inset) through the serially connected ion channel array (as a discrete resistor). The neuron forms a simple \textit{serially }(\textit{not in parallel}, as assumed by Hodgkin and Huxley~\cite{HodgkinHuxley:1952} and mistakenly claimed by neurophysiology) connected $RC$ oscillator.
For the physical and mathematical description, see section~4 in~\cite{VeghTechnomorphBiology:2025}; for its algorithmic specifics~\cite{VeghNeuronAlgorithms:2025}, for further details~\cite{VeghDANCES:2026}.
Notice that the difference between the rush-in and the resultant gradients
emphasizes the role of the finite resources: without the \gls{AIS},
no turn-back of the resultant gradient would occur.

As is well known from the theory of electrical circuits, the output
voltage measured on the output resistor is as follows:

\begin{equation}
	V_{out}^{Differentiator}=RC\frac{dV_{in}}{dt}\label{eq:RC_Circuit_Output}
\end{equation}

\noindent where the input voltage is as follows:

\begin{equation}
	\frac{d}{dt}V_{in}=\sum\frac{d}{dt}V_{IN}^{Component}-\frac{d}{dt}V_{OUT}^{AIS}\label{eq:RC_Circuit_Input}
\end{equation}

\noindent that is, the (temporally gated) sum of the input gradient
that the currents generate, plus the gradient of the output current
through the \gls{AIS}~\citep{AISStructureReview:2018,AIS_Updated_Viewpoint:2018}.
The latter term can be described as follows:
\begin{equation}
	\frac{d}{dt}V_{OUT}^{AIS}={\frac{1}{C}}\frac{V_{internal}-V_{external}}{R_{AIS}}\label{eq:AIS_Voltage}
\end{equation}

The fundamental difference between the parallel and serial $RC$ circuits is that \textit{the parallel circuit cannot produce output voltage with opposite sign (so it requires an ad-hoc hypothesis of a delayed current in the opposite direction), while the serial one changes the sign of its output
	for rising and falling edges of the input voltage}. 
As discussed, the capacitive current, which by definition changes its direction and so 
generates an opposite voltage on the resistor represented by the \gls{AIS}, perfectly describes the so-called "hyperpolarization". It is not the effect of a $K^+$ current through the membrane: the resting ion channels do not have sufficient conductance.  
Not knowing about the \gls{AIS} leads to assuming that the resting ion channels work also as transient channels (in other words, assuming a parallel $RC$ circuit instead of the serial one, furthermore, that
the output current flows through the membrane instead of the axon). Some $K^+$ current certainly exists; for the magnitude of currents during \gls{AP}, see~\cite{NeuralEnergyConsumption:2017}.

The shape of the output waveform depends on the ratio of the pulse width to the $RC$ time constant. When $RC$ is much larger than the pulse width, the output waveform resembles the input signal, even with a square wave input. (In the case of the neuronal oscillator, the shape of the front side of the spike is similar to the one derived for the rush-in current, while the back side is very prolonged.)

\subsection{The elastic model\label{sec:Elastic}}

\begin{figure}
	\includegraphics[scale=1.5]{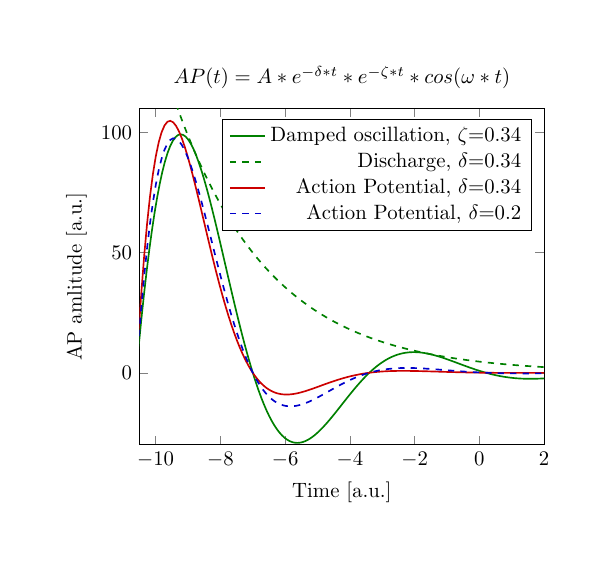}
	
	\caption{Describing the generation of an \gls{AP}
		as the superposition of an elastic vibration 
		of the membrane and an electrical discharge process.
		\label{fig:DampedOscillation}}
	
\end{figure}

The rush-in of $Na^+$ ions results in electric, chemical, and elastic gradients, see~\cite{VeghUnifiedCrossdisciplinaryModel:2026}. The final result (shown in Fig.~\ref{fig:DampedOscillation}) is that the membrane vibrates the 
electrolyte (with the cca. $10^{-3}$ part ions inside).
Moreover, the electrostatic repulsion and the concentration gradient of ions provide an offset voltage that triggers an electrical discharge. 
The two processes are superimposed.

The sudden rush-in of
$Na^+$ ions has (at least) a double effect. 
One effect is that they press the surface of the membrane that
elastically increases its diameter~\cite{NeuronalDeformation:2020}.
Due to their mutual repulsion,
the ions experience a strong electrical force toward the membrane. It represents an impulse $J=F\,\Delta t$ [N\,s], that enables estimating the time and energy of the 
change the rush-in causes and explains that the energy needed to 
generate an \gls{AP} is suddenly produced electrically, 
and stored temporarily as elastic energy (the neuron produces the required energy in the background, when \gls{ATP} produces ions by hydrolysis under the effect of the electrical field near to membrane); see~\cite{VeghUnifiedCrossdisciplinaryModel:2026}. 
The case is practically identical to
the deformation of an elastic plate induced by the hydrostatic pressure of a water column. Initially, the aluminum plate is suddenly subjected to hydrostatic pressure, and it reaches equilibrium after initial oscillations. The diagram line of the process is shown in the figure
\cite{HydrostaticWaterColumn:2021}, and the method of simulation is discussed in \cite{Hydrostatic_SPHinXsys:2021}. Compare the diagram line to the
gradient in Fig.~\ref{fig:PhysicalProcessesMembrane}, middle inset.

The elastic force is described by the  
\[
F(t) = F_{max}^{elastic}\times e^{-\zeta*t}\times cos(\omega\times t)
\]
\noindent 
formula. The classical model entirely neglects this elastic gradient. However, it is almost four orders of magnitude higher than the electrical and thermodynamic gradients (see Appendix~\ref{sec:Forces}), explaining why the damped elastic oscillation almost perfectly describes 
the time course of the \gls{AP}. However, the elastic force oscillates and decays exponentially, so the dominance of 
the elastic term changes with time.
The electrical model can perfectly describe \gls{AP} since 
the pressure and the electric field are proportional to the number of ions inside a volume.

The other effect is that a large amount of positive charges
appears in the neuron, significantly increasing the membrane's potential. The axon remains on the resting potential, so the potential gradient drives a current.
A discharge process takes place where the amount of charge
inside the neuron exponentially decreases 
\[
F_{electric}(t) = F^{max}_{electric}\times e^{-\delta*t}
\]

The two effects act at the same time, that is
\[
F_{resulting}(t) = F^{max}_{resulting}\times e^{-\delta*t}  \times e^{-\zeta*t}  \times cos(\omega\times t)
\]
(the two exponential factors have physically different reasons,
but they can be contracted mathematically), resulting
an oscillation shown in Fig.~\ref{fig:DampedOscillation}.
This model combines the effects of two distinct disciplines,
but may not be perfect, because of the dual role of ions:
the electric effect changes the oscillating mass and the oscillation changes the amount of charge. However, it 
suggests a way to synthesize the two effects, and explain 
a till now unexplained thermodynamical observation.

Furthermore, the  elastic model explains the experienced constant intensity
of \gls{AP} (without the ad-hoc hypothesis of cooperating
ion channels in the axon's wall). The vibrating electrolyte 
behaves as a 'soliton'~\cite{SolitonPropagation:2005}, that is, transfers the pressure wave with a minimum material transport and intensity loss, and the dissociated ions inside the vibrated electrolyte generate the potential field observed as \gls{AP}.

\subsection{Heat emission/absorption\label{sec:HeatAbsorption}}

The early experimental observation~\cite{HeatProductionNeuron:1958}, that positive and negative heat production is associated with a nerve impulse, suggested a thermodynamic model. 
In the framework of the classical
theory, Ohmic currents ﬂow through resistors that dissipate heat due to
friction, no matter in which direction ($W=I^2\times R$) the ion currents ﬂow.
The existence of "cooling", alone, undermines the credibility of the classic theory~\cite{HEIMBURGReversibleHeatProduction:2021}.
The issue here is that the dissipation was attributed to the "leakage current."  
As Hodgkin wrote: "Hill and his colleagues found~\cite{HeatProductionNeuron:1958} that an initial phase of heat liberation [of the action potential] was followed by one of
heat absorption. [...] a net cooling on open-circuit was totally unexpected
and has so far received no satisfactory explanation."~\cite{HodgkinConduction:1964}
"All authors came to similar conclusions: during
the action potential, no signiﬁcant net heat is produced. Transient heat releases are mostly reabsorbed in the second phase of the action potential. \dots The ﬁnding of a reversible heat release during the action potential of
nerves is a striking and very fundamental fact. It is inconsistent with the
\gls{HH} model. The physics underlying the nervous impulse
must rather be based on reversible processes".~\cite{ThermalBiophysics:2007} 
At that time it was commonly known that compressing a
medium produces heating and extending it causes cooling.
Although modeling neuron as a thermodynamic engine 
that is heated and cooled, and produced no signiﬁcant net heat, would be trivial, no such model came to light.
The probable reason is that no reasonable idea
shined up that could attach the observed heat absorbtion
with the observed evident signs of the electrical operations.

Heat liberation and absorption are direct experimental evidence of the "slow current" and the mechanisms described above.
The lack of leakage current
not only validates our model, but also solves the long-standing mystery of reversible heat release during the \gls{AP}.
Pressure changes accompany the electrical process and naturally account for the observed temperature changes.
The energy consumption of the nervous impulse underpins that the complex physical process comprises mostly reversible disciplinary processes:
most of the energy of \gls{AP} is stored as reversible elastic potential energy. Hence, its dissipation is also negligible: "no signiﬁcant net heat is produced"~\cite{ThermalBiophysics:2007}. 

The shocking conclusion in~\cite{MechanicalBrainPulses:2018}, that "brain cells communicate
with mechanical pulses, not electrical signals", more precisely sounds
that the elastic membrane generates a pressure wave in the volume of the 
membrane, and all components having mass (the electrolyte) vibrate "in place". The charged components (the ions, a $10^{-3}$ part of the electrolyte) repel each other (as long as there is excess charge on the membrane) toward the \gls{AIS} and along the membrane.
Actually, the disciplines of thermodynamics and electricity cooperate in both producing and transmitting \gls{AP}. Neither of them can do that alone. No compromise is needed.
The nervous pulse transmission over the axons 
also must be revisited. The classical mechanism based on coordinated action
of ion channels is undoubtedly wrong. Among others, the physical conditions of applicability of the telegrapher's equations are not fulfilled, plus (see Appendix~\ref{sec:Forces}) the electrical forces are orders of magnitude weaker than the elastic ones. The enormous pressure, the incompressible electrical fluid,
and the repulsion of the delivered particles suggest 
an alternative and more reasonable mechanism.
The resultant of the electrical and thermodynamic, and first of all, the mechanical force generated by the elastic membrane, does the task.
No protein mechanisms and cooperating ion channels are needed.

\begin{figure}
	\includegraphics[width=.9\columnwidth]{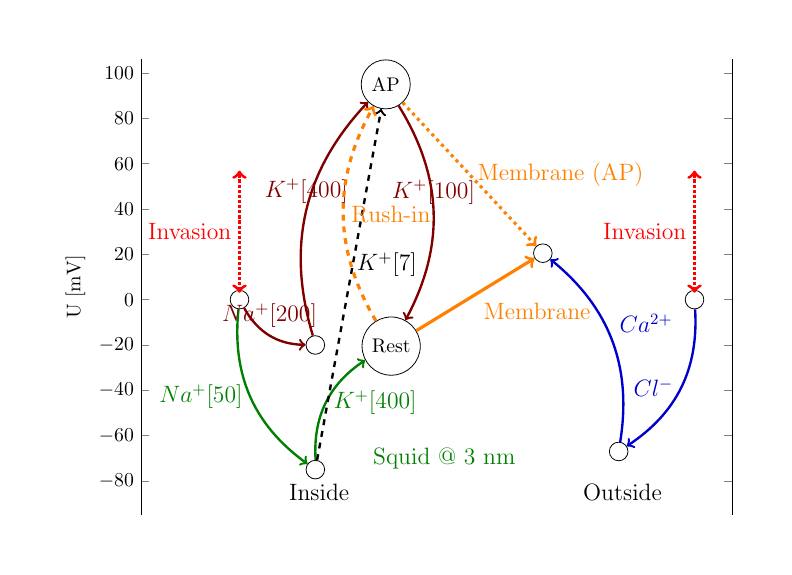}
	\caption{The figure is supplemented with the unbent black line showing that the clamping condition forces the neuron
		to find a new balanced state by compensating the clamping voltage
		by setting a concentration [Na50,K7].
		For comparison, the case of natural \gls{AP} is also shown.
		The mechanism of producing \gls{AP} (using numbers referring to the case of squid shown in Table~1 in~\cite{VeghUnifiedCrossdisciplinaryModel:2026}). In the resting state, the inside thermal potentials follow the green paths.
		When the rush-in of $Na^+$ ions suddenly increases the internal concentration to $[200]$, the potential on the internal side of the membrane increases to $+90\ mV$. The neuron must decrease its concentrations to restore the resting potential, mainly by releasing an action potential through the \gls{AIS}.
		The clamping fixes the neuron's membrane voltage at \gls{AP},
		so the neuron can operate only with the concentrations.
		\label{fig:RestingPotential5}
	}
\end{figure}

\section{Measuring the outward $K^+$ current by clamping\label{sec:OutwardK}}

It has been a mystery since the  \gls{HH} model was created: how \gls{AP}
can also take negative values when only
$Na^+$ influx was seen when \gls{AP} was formed? 
It looks like that neither in the electrical, nor in the mechanical/thermodynamic/elastic view (which, by the way, are two sides of the same coin), exists a place for the outward $K^+$ current
through the membrane.
Since the hypothesized process of $K^+$ outflow was too fast, \gls{HH} used a voltage clamping method for its experimental determination~\cite{HodgkinHuxleyCurrent:1952}. The essence of their procedure was that they
imitated crossing of the membrane voltage threshold (reaching the state "AP"), then kept the membrane's potential above the resting level:
"after completion of the quick pulse through the membrane
capacity",  
"the potential
difference across the membrane is suddenly changed from its resting
value, and held at the new level by a feedback circuit ('voltage clamp' procedure)", and measured the time dependence of the current in this state.

\subsection[The formal description of the process]
{The formal description of the process\label{sec:Fallacies-OutwardKPhysicsFormal}}

Figure \ref{fig:RestingPotential5} shows the changes that occur in a neuron that is displaced from its equilibrium state.
These processes are naturally slow; their role in the
transient process is discussed elsewhere~\cite{VeghUnifiedCrossdisciplinaryModel:2026},
here we only illustrate our statements regarding the $K^+$ current.
Our statement is essentially based on Kirchhoff's Voltage Law,
according to which the sum of the voltages in a closed circuit is zero.
In our case, the circuit extends from the inside of the neuron to the external environment.
If, for any reason, the external and internal potential levels are not the same, a current flows until they equalize (current carries charge and causes a potential change; it is not an ideal voltage generator).
Of course, the equalization is not
instantaneous; the current is time-dependent.

On both the left and right sides of the figure, the resting potential level is
at the zero value of the vertical scale. In the resting state (without invasion),
in the intracellular segment
the [Na50,K400] concentrations are present inside the cell. Using these values, we reach the "Rest"
state along the green arrows (based on the Nernst voltages, see Table~1 in~\cite{VeghUnifiedCrossdisciplinaryModel:2026}), which is located at a height of \SI{-21}{\milli\volt}.
On the right side, in the extracellular segment, starting from the point at height 0, we arrive at a point at \SI{21}{\milli\volt} along the blue arrows in the same way. The potential difference between the two mentioned points is \SI{42.5}{\milli\volt}, which \gls{HH} measured. It is a correct value,
but it is the result of an electrostatic charging and not of a voltage drop due to the "leakage current" they assumed.
\index{leakage current}
The voltage across the membrane (represented by the solid orange arrow) is directed upward, preventing positive ions in the intracellular space from moving toward the extracellular space and preventing negative ions in the extracellular space from moving toward the intracellular space. The $Ca^{2+}$ ions in the extracellular segment, however, are permeable and require continuous pumping. There is no potential difference between the outside and inside, so no (significant) current flows.

If, for any reason, the membrane voltage suddenly rises above the threshold value, due to the extremely fast $Na^+$ influx,
the membrane potential increases by about \SI{100}{\milli\volt} (a [Na200,K400] state occurs, see the brown arrows), i.e., the neuron jumps into the AP bubble.
The orange arrow that was pointing upwards now points downwards as a broken line: the outflow of positive ions (both $K^+$ and $Na^+$) is given a free path, but the sloping path for $Ca^{2+}$ ions is eliminated. However, the free ions in the surface layer of the membrane are the vast majority of the $Na^+$ ions that have just arrived, so the ion current through the \gls{AIS} mainly consists of such ions.
A significant increase in the membrane voltage leads to a significant increase in the potential of the
inner space of the neuron.
A current is initiated that tries to restore the initial voltage and concentration relationships.
The extracellular environment of the membrane is extremely stable (the concentrations do not change),
but inside the cell the voltage increases to the same extent, in the direction of "Invasion".
The membrane potential has changed significantly due to the change in the $Na^+$ concentration, so the cell then moves them towards equilibrium.

In the case of natural \gls{AP}, there is no obstacle for the neuron
to restore its previous state by changing any parameter, so the system concentrations
move towards the state [Na50,K400]. This can be
easily achieved by trying to make Na[50] from Na[200], while $K^+$ is still/already at the appropriate concentration.
That is, it releases (mainly) $Na^+$ ions through the \gls{AIS} until the potential difference drops to zero.
Reducing the $Na^+$ concentration also reduces the membrane
voltage, i.e., after the release of the excess ions (admitted during the $Na^+$ rush-in) the neuron returns to its initial state;
a cycle takes place.

"Clamping" fixes the membrane voltage at a value different
from the one the electric/ thermodynamic backbone defines, meaning that the neuron cannot restore its initial
resting state. Instead, it has to find a new balance by changing the concentrations.
After the artificial \gls{AP}, clamping (using negative feedback) keeps the membrane potential at a fixed value, so the neuron can only produce a new balance by changing the ion concentrations: a current flows until the Nernst voltages of the $Na^+$ and $K^+$ concentrations can balance the clamping voltage.
From this [Na50,K400] state, the neuron has to adjust the potential so that the sum of the Nernst voltages plus the clamping voltage is zero (another balance has to be set!), while the voltage is fixed.
To do this, it must reduce the $K^+$ concentration, while the $Na^+$ concentration is already at a good value (the opposite of the previous case). And not by a small amount: [K7] only causes a voltage change equal to the clamping 100 mV imposed on the cell.
This voltage is given by the green bent $Na^+[50]$ and the blue straight
$K^+[7]$ path. The neuron can achieve this [Na50,K7] concentration
by starting an intensive $K^+$ release.

Another important difference compared to the natural case is that there is no electrostatic acceleration across the membrane (which is why the $Na^+$ influx stopped),
only a thermodynamic suction effect outwards due to clamping.
Indeed, $K^+$ ions attempt to flow out through the membrane, but
they cannot exit until the electric potential 
in the layer near to the membrane drops
below the equivalent thermodynamic driving force of the concentration.
Moreover, the nearby $K^+$ ions arrive sooner, but they leave a gap (reduced concentration) behind them, where the nearby-neighbor $K^+$ electrodiffuses. This is why the physiologist sees a "delayed current".
It is a strange game of nature that the resultant of the thermodynamic and the electric  (plus the elastic)  gradients moves, simultaneously, the $Na+$ ions parallel to the membrane surface, toward the \gls{AIS}, and
the $K^+$ ions perpendicular to the surface toward the membrane,
approximately until the peak of the action potential voltage is reached.
The $Na^+$ ions start to leave through the \gls{AIS}, while the "excessive" $K^+$ ions leave through the membrane.
\textit{The two currents do not flow in the same path, so they cannot
	interfere as currents in opposite directions and cannot cause hyperpolarization.}
\index{hyperpolarization}
Hyperpolarization is caused by the reversal of the $Na^+$ current, which can be interpreted
in the electrical sense as the capacitive current of the capacitor in the serial $RC$ circuit, and
in the mechanical sense as the negative period of the elastic membrane oscillation.

\begin{figure}
	\includegraphics[width=.8\columnwidth]{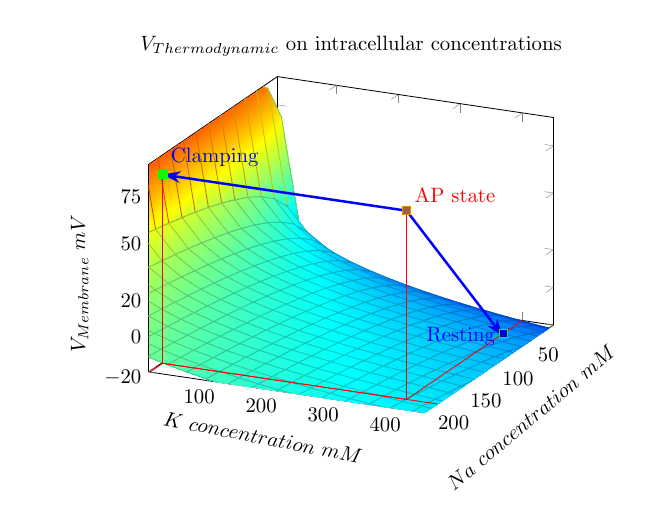}
	\caption{Artificial and natural degradation of the AP state.
		In the case of natural degradation, the neuron
		has to get rid of the (unnecessary) $Na^+$ ions;
		by releasing them, it returns to the initial state;
		cycle takes place.
		Clamping keeps the membrane voltage fixed,
		so the neuron can only establish a new equilibrium situation by changing the intracellular concentrations
		which, due to the very reduced $K^+$ concentration, is only possible with an intense
		$K^+$ release.
		\label{fig:ClampingKCurrent}
	}
\end{figure}

\textit{That is, the $K^+$ current is an artifact caused by the clamping, which is caused by the 'clamping' measurement conditions, and does not occur in the neuron under natural conditions.}

In the classic \gls{HH} experiment, clamping means that the left-hand "Invasion" suddenly raises the membrane voltage to the peak value measured at \gls{AP}, while the concentrations remain unchanged, so the voltage of the left-hand segment will also be higher.
The current only stops when the difference in the Nernst voltages of the inner segments equalizes
the clamping voltage. Due to the slowness of charge carriers, the process is not instantaneous.

Fig.~\ref{fig:ClampingKCurrent} shows a pictorial summary of the process and illustrates the native and 
clamped experimental conditions, showing that those targets are physically different states.
The points on the surface give the Nernst voltage acting on the ions, for different combinations of concentrations. The points on the surface are in equilibrium. The onset of the \gls{AP} moves the point that was in equilibrium until then well outside the surface. This non-equilibrium state is the same for both the natural and artificial excitation. The neuron can find the new equilibrium position without the clamping force by restoring the parameters to their previous state. In the case of clamping, it must set new concentrations at which the sum of the Nernst potentials can offset the clamping voltage. In both cases, a relaxation process occurs, which, in the case of the native process, is measured physiologically as \gls{AP}.

% Compiled 26.07.10
\subsection[The physics of the process]
{The physics of the process\label{sec:Fallacies-OutwardKPhysics}}

The essential functioning of the cell largely occurs in layers a few nanometers thick near the cell membrane, where concentration and voltage gradients (and consequently material flow rates) can differ by up to several orders of magnitude for a few dozen microseconds. 
Furthermore, the slow material flow creates potential and concentration changes that vary from moment to moment, and whose effects can differ significantly across layers due to the chemical quality of the ions. 
For this reason, the description of the events is unique in each case, for each layer and each time period; it is not uncommon for scientific disciplines to change when describing individual stages, or for them to be combined differently.

The two ends of the ion channels embedded in the membrane have different potentials and concentrations, so that the ions passing through acquire significant energy and speed. However, this gradient disappears when they exit the channel, so that the ions "stop" in frustration.The passage continues until the passing ions increase the electrochemical potential on the other side to such an extent that the gradient disappears. Inside the electrolyte, the other, orders-of-magnitude smaller gradients continue to operate, of course, but can only create very slow material transport. The changes in the "bulk" part of the electrolyte tend towards slow equalization (different rules remain in effect near the boundary membrane). For this reason, a slow diffusion effect is also associated with all processes.

When $Na^+$ ions penetrate the intracellular segment, they do so under the influence of a significant electrochemical gradient.  In this way, immediately after the passage (lasting a few nanoseconds), the ions create a $Na^+$-rich layer, from which the ions' charge (partly by direct collisions) knocks out the ions there, mainly $K^+$.
Basically, a $Na^+$ layer forms directly beneath the intracellular part of the membrane, thereby increasing the voltage across the capacitor plates. Since the \gls{AIS} at the end of the membrane acts as a sink for the ion current, an immediate $Na^+$ current is initiated, as discussed in the electrical model.

$Na^+$ ions move along the membrane surface due to the high electrical gradient through the high-conductivity \gls{AIS}, but small amounts of $Na^+$ and $K^+$ ions apparently diffuse into the adjacent layer. The ion concentration and potential of the layer decrease as the outflow current increases. Due to the slowness of the ion current, the ions only reach the \gls{AIS} in tens of microseconds.

The membrane is initially closed to positive ions (after the $Na^+$ ions have passed through), and the $Na^+$ layer represents an additional potential barrier. However, over time, the decreasing $Na^+$ concentration due to the current in the layer that causes the \gls{AP} increasingly allows $K^+$ ions to move towards the membrane and the experienced $K^+$ current to start, which is only possible after a few dozen microseconds.

"Delayed $K^+$ channels" therefore means ion channels that structurally do not distinguish between ions, and the delay is caused by transient processes taking place in the ion layers. These channels indeed release $K^+$ ions, which only start when the clamping voltage is applied to the neuron (i.e., they create a non-equilibrium state of the neuron segment), and only $K^+$ ions because the thermodynamic gradient is favorable only for them.

\gls{HH} found that the \gls{AP} can be decomposed into two currents:
"the ionic current during a depolarization consists of two more or less independent components in parallel, an early transient phase of current carried by sodium ions, and a delayed long-lasting phase of current carried by potassium ions". Really, the current component arising from Na ions appears quickly, the component consisting of $K^+$ ions remains delayed and persistent.
Their conclusion, however, is more than surprising.  According to that "explanation", the neurons are disposable devices: after emitting a single \gls{AP} once in a while, the neuron emits a short-lasting $Na^+$current and (according to their figures 5 and 6) an infinitely long-lasting $K^+$ current, contrary to experience.
In other words, for \SI{100}{\pico\ampere} (that is \SI{1e-10}{\ampere}) saturation current and \SI{1e11}{neuron} it would mean
10~amper permanent current in the brain.
Furthermore, it would keep the resting potential above 
the value of the resting potential. The difference 
would be the extra voltage of the clamping; that is,
it would disappear at zero clamping voltage.
In other words, the extra voltage and the  $K^+$ current disappear in the absence of clamping, as observed.

So, this is where the urban legend that "$K^+$ flows out through the membrane in the second phase of \gls{AP}, with a significant delay compared to the $Na^+$ influx" comes from.
It is a real measurement; it is $K^+$ ions, but \gls{HH}did not measure what they thought they would.
So there is no question of specific $Na^+$ and $K^+$ channels (synchronized with each other): in both cases, the concentration and potential relationships decide what ions flow and in which direction.
In our reading, the current lasts as long as the voltage invasion caused by clamping persists or until the $K^+$ content of the neuron's intracellular segment is depleted.

\section{Summary}

A ghost component, a $K^+$ current, misled neurophysiology. 
The starting point, taken decades before discovering \gls{AIS},
from where the recent electrical theory started, was wrong: 
\gls{AIS} is a major player that has enormously changed the rules of the game. Unfortunately, several ad-hoc hypotheses were introduced
into the theory of neuronal electrical operation before discovering \gls{AIS}. The idea of the phantom current was introduced into the theory because the that-time resolution of microscopes did not enable to construct a theory that includes \gls{AIS}. The incomplete understanding of clamping and neuronal operation resulted in an artifact that 
enabled to measure a genuine $K^+$ current, but in a state
different from the native \gls{AP} current. These two mistakes confused research in many aspects. We succeeded to identify the fictive $K^+$ current as the correct model's capacitive current. 

\begin{appendix}
	\section{Forces exerting on ions\label{sec:Forces}}
We give some plausible estimations for the forces
exerting on ions. Although they may not be exact numbers,
they are surely in the right order of magnitude, and they 
underpin quantitatively the conclusions above.

 By using that in an electric field $\vec F = q \vec E$, one can calculate the force exerted on a charge, that enables one to calculate also the particle's impulse, pressure, and energy. For example, in the case of $Na^+$ rush-in, the electric force field is $10^7$ [V/m].
Since($[N/C]==[V/m]$),
the value of the force exerted on a single ion, accelerating it across the ion channel, is
\begin{equation}
	F_{Na^+}= 10^7  [N/C] \times 1.602* 10^{-19} [C] = 1.6*10^{-12}\ [N] \label{eq:UnitForce}
\end{equation}
\noindent
One can estimate how the pressure of the neural cell increases due to the rush-in changes at the beginning of the
\gls{AP}. 
As evidence shows, the local potential at the internal surface of the membrane is in the range of \SI{100}{\milli\volt} in the resting state and increases by $\Delta U=$ \SI{100}{\milli\volt}  in the transition state.
% This increase means a change in the force acting on an ion (see Eq.~(\ref{eq:UnitForce})) by  $1.6*10^{-12}\ [N]$.
Of course, the membrane exerts a counterforce of the same size
on the ion. In the moment after the rush-in, $F_{max}^{elastic}$ casuses a pressure change  Eq.~(\ref{eq:CellPressureChange}).
When we assume $10^7$ rush-in ions and unchanged cell size, the total force acting on the membrane increases by $1.6*10^{-5}\ [N]$.
This change in force means on the neuron's $8*10^{-9}\ [m^2]$ surface  a pressure change 
\begin{equation}
	\Delta P = \frac{1.6*10^{-5}\ [N]}{8*10^{-9}\ [m^2]}
	= 2*10^{3}\quad \biggl[\frac{N}{m^2}\biggr]
	\label{eq:CellPressureChange}
\end{equation}
This pressure on a $10\times10~\mu m$ tooltip is converted to a force
$2*10^{3}*100*10^{-12} = 200~pN$ force. The measured force value~\cite{MechanicalPropertiesNerves:2025} is about $600~pN$,
so our estimation is in the correct range.

The \SI{100}{\milli\volt} potential difference on the \SI{50}{\micro\meter} length of the \gls{AIS} generates a field
\SI{2e3}{\volt\per\meter}
that exerts on the ions
\begin{equation}
	F_{Na^+}= 2*10^3  [N/C] \times 1.602* 10^{-19} [C] = 3.2*10^{-16}\ [N] \label{eq:ElasticForce}
\end{equation}
maximum force. 

\end{appendix}

\section*{Competing interests}
The authors declare that they have no competing interests

\subsection*{Funding}
The research did nor receive any support.

%%%\bibliographystyle{elsarticle-num} 
%\bibliography{/home/jvegh/REPO/LaTeX/CommonBibliography,/home/jvegh/REPO/LaTeX/CommonNeuronalBibliography,/home/jvegh/REPO/LaTeX/CommonPrivateBibliography,/home/jvegh/REPO/LaTeX/CommonAIBibliography}

\end{document}